\documentclass{article}
\usepackage{spconf,amsmath,graphicx}
\PassOptionsToPackage{hyphens}{url}\usepackage{hyperref}

\usepackage{xcolor}

\newcommand\figref[1]{Fig.~\ref{#1}}
\newcommand\tabref[1]{Table~\ref{#1}}
\newcommand\secref[1]{Section~\ref{#1}}

\newcommand{\sedit}[1]{\textcolor{black}{#1}}

\title{Period VITS: Variational Inference with Explicit Pitch Modeling\\ for End-to-end Emotional Speech Synthesis}
\makeatletter
\def\@name{
  \emph{Yuma Shirahata}$^{1}$,
  \emph{Ryuichi Yamamoto}$^{1}$,
  \emph{Eunwoo Song}$^{2}$,\\
  \emph{Ryo Terashima}$^{1}$,
  \emph{Jae-Min Kim}$^{2}$,
  \emph{Kentaro Tachibana}$^{1}$\\
}
\makeatother
\address{
  $^{1}$LINE Corp., Tokyo, Japan,\\
  $^{2}$NAVER Corp., Seongnam, Korea
}
\begin{document}
\maketitle
\fontsize{9.0}{10.4}\selectfont

\begin{abstract}
Several fully end-to-end text-to-speech (TTS) models have been proposed that have shown better performance compared to cascade models (i.e., training acoustic and vocoder models separately).
However, they often generate unstable pitch contour with audible artifacts when the dataset contains emotional attributes, i.e., large diversity of pronunciation and prosody.
To address this problem, we propose Period VITS, a novel end-to-end TTS model that incorporates an explicit \textit{periodicity generator}.
In the proposed method, we introduce a \textit{frame pitch predictor} that predicts prosodic features, such as pitch and voicing flags,  from the input text.
From these features, the proposed periodicity generator produces a sample-level sinusoidal source that enables the waveform decoder to accurately reproduce the pitch.
Finally, the entire model is jointly optimized in an end-to-end manner with variational inference and adversarial objectives.
As a result, the decoder becomes capable of generating more stable, expressive, and natural output waveforms.
The experimental results showed that the proposed model significantly outperforms baseline models in terms of naturalness, with improved pitch stability in the generated samples.
\end{abstract}
\section{Introduction}
\label{sec:intro}
Text-to-speech (TTS) has recently had a significant impact due to the rapid advancement of deep neural network-based approaches~\cite{tan2021survey}.
In most previous studies, TTS models were built as a cascade architecture of two separate models---an acoustic model that generates pre-defined acoustic features (e.g. mel-spectrogram) from text~\cite{ren2019fastspeech,wangIS2017tacotron} and a vocoder that synthesizes waveform from the acoustic feature~\cite{oord2016wavenet,yamamoto2020parallel,kong2020hifi}. Although these cascade models were able to generate speech reasonably well, they typically suffered from an error deriving from the use of pre-defined features and separated optimization for the two independent models. Sequential training or fine-tuning can mitigate the quality degradation~\cite{shen2018natural}, but the training procedure is complicated.

To address this problem, several works have investigated the use of fully end-to-end architecture that jointly optimizes the acoustic and vocoding models\footnote{We refer to text-to-wave model as end-to-end TTS throughout this paper. Note that the term is not used for cascade type models.}~\cite{ren2020fastspeech,donahue2020end,kim2021conditional,lim2022jets}.
One of the most successful works is VITS~\cite{kim2021conditional}, which adopts a variational autoencoder (VAE)~\cite{kingma2013auto} with the augmented prior distribution by normalizing flows~\cite{rezende2015variational}.
The VAE is used to acquire the trainable latent acoustic features from waveforms, whereas the normalizing flows are used to make the hidden text representation as powerful as the latent features.

However, we found that although VITS generates natural-sounding speech when trained with a reading style dataset, its performance is limited when applied to more challenging tasks, such as emotional speech synthesis, where the dataset has significant diversity in terms of pronunciation and prosody.
Specifically, the model generates less intelligible voices with unstable pitch contour.
Although the intelligibility problem could be addressed by expanding the phoneme-level parameters of prior distribution to frame-level parameters~\cite{zhang2022visinger}, it is still a challenge to generate accurate pitch information due to the architectural limitation of the non-autoregressive vocoders~\cite{morrison2021chunked}.

To tackle this, we propose Period VITS, a novel TTS system that explicitly provides \textit{sample-level and pitch-dependent periodicity} when generating the target waveform.
In particular, the proposed model consists of two main modules termed the prior encoder and the 
waveform decoder (hereinafter simply called ``decoder").
On the prior encoder side, we employ a frame prior network with a frame pitch predictor that can simultaneously generate the parameters of the prior distribution and prosodic features in every frame.
Note that
the parameters are used to learn expressive prior distribution with normalizing flows and
the prosodic features such as the pitch and the voicing flags are used to produce the sample-level sinusoidal source signal.
On the decoder side, this periodic source is fed to every up-sampled representation in the HiFi-GAN-based vocoder~\cite{kong2020hifi}, \sedit{to guarantee pitch stability in the target waveform.}
Note that the training process optimizes the entire model in the end-to-end \sedit{scheme} from the variational inference point of view.

Several works have addressed similar problems by focusing on the periodicity of speech signals in the vocoder context ~\cite{hono2021periodnet,hwang2021high,yoneyama22_interspeech}.
The proposed model differs in that the conventional methods have used pre-defined acoustic features and only optimized the vocoder part separately.
In contrast, the proposed architecture has the benefit of end-to-end training and obtains optimal latent acoustic features guided by the auxiliary pitch information.
In addition, another prior work has tackled the pitch stability problem by adopting a chunked autoregressive architecture in the vocoder~\cite{morrison2021chunked}.
Unlike \sedit{that method, our proposal} can generate waveforms in much faster speed, thanks to the fully non-autoregressive model architecture.

The experimental results show that the proposed model performed significantly better than all the baseline models including end-to-end and cascade models in terms of naturalness in the  multi-speaker emotional TTS task. Moreover, the proposed model achieved comparable scores to the recordings for neutral and sad style with no statistically significant difference.

\section{PROPOSED METHODS}
\label{sec:proposed}
\begin{figure}[t]
    \centering
    \includegraphics[width=8cm]{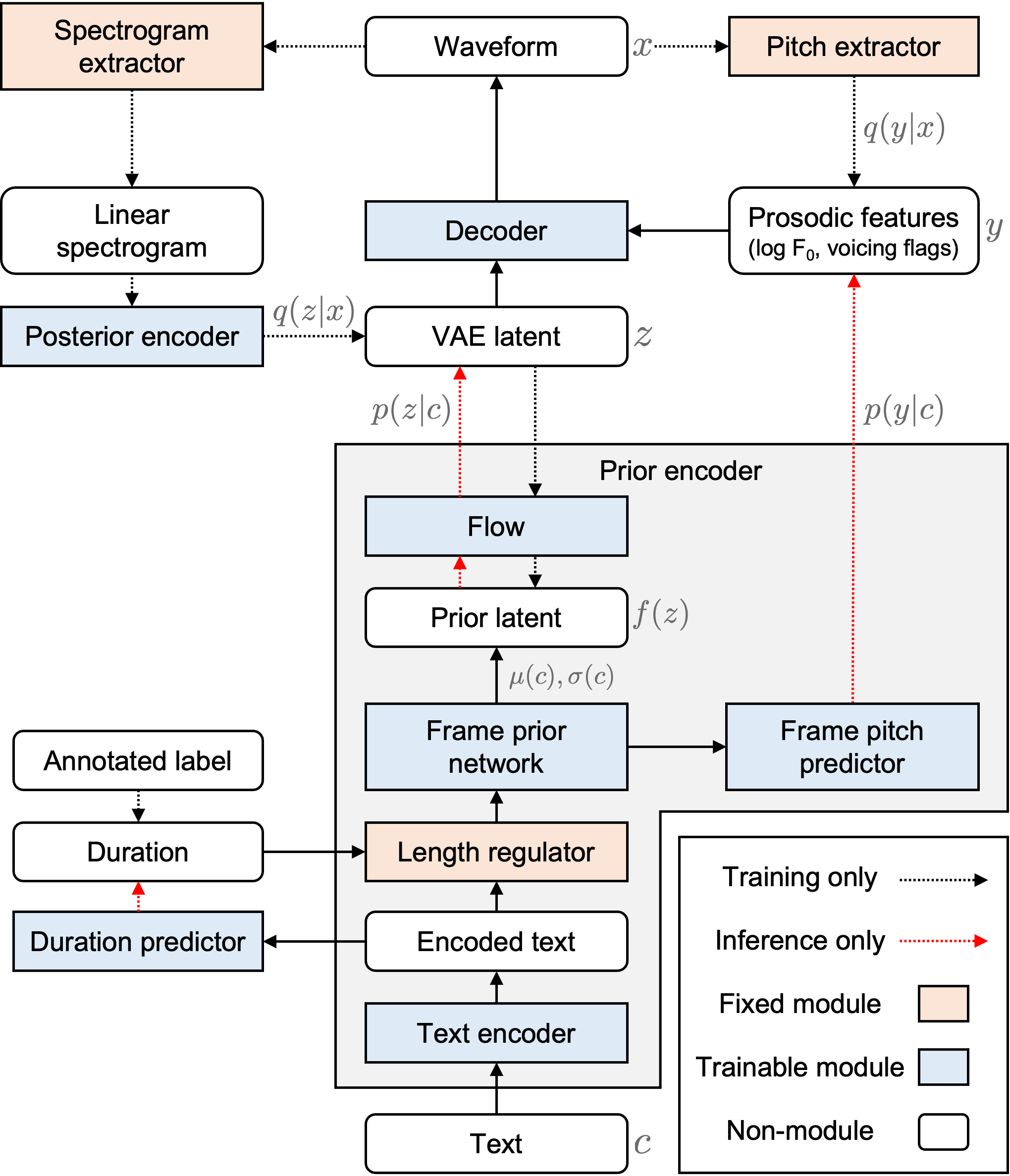}
    \vspace{-2mm}
    \caption{
    \fontsize{9.0}{10.6}\selectfont
    The architecture of the proposed model. During training, a waveform is reconstructed from the latent feature from the posterior distribution $q(z|x)$ and the extracted pitch distribution $q(y|x)$. During inference, the latent feature from the prior distribution $p(z|c)$ and the predicted pitch distribution $p(y|c)$ is used alternatively.}
    \label{fig:phgvits}
    \vspace{-2mm}
\end{figure}

\subsection{Overview}
\label{subsec:model_overview}
The overall model architecture is shown in \figref{fig:phgvits}. Inspired by VITS~\cite{kim2021conditional}, we adopt a VAE whose prior distribution is conditioned on text. The VAE is composed of a posterior encoder and a decoder, whereas the prior distribution is modeled by a prior encoder.

Specifically, the posterior encoder converts the input linear spectrogram to latent acoustic features, while the prior encoder transforms the text into the latent features. The decoder is responsible for reconstructing waveforms from the learned latent features.

In addition, we introduce a latent variable $y$ to represent the prosodic feature as a separated source from the VAE latent variable $z$ to explicitly model the pitch information of the generated speech.
The proposed model is trained to maximize the log-likelihood of waveform $x$ given text $c$. However, as it is intractable, we optimize a lower bound on the marginal likelihood, as follows~\cite{kingma2013auto}:
\begin{align}
    &\log p(x|c) = \log \iint p(x,z,y|c)dz dy \notag\\
    &\geq \iint q(z,y|x)\log\frac{p(x,z,y|c)}{q(z,y|x)} dz dy \notag\\
    &= \iint q(z|x)q(y|x)\log\frac{p(x|z,y)p(z|c)p(y|c)}{q(z|x)q(y|x)} dz dy \notag\\
    &= E_{q(z,y|x)}[\log p(x|z,y)] - D_{KL}(q(z|x)||p(z|c)) \notag\\
    &- D_{KL}(q(y|x)||p(y|c)), \label{eq:elbo}
\end{align}
where $p$ denotes the \sedit{generative model} distribution, $q$ denotes an approximate posterior distribution for $z$ and $y$, $E$ denotes the expectation operator, and $D_{KL}$ represents the Kullback--Leibler (KL) divergence.
In addition, we assume $z$ and $y$ are conditionally independent given $x$\footnote{This assumption is reasonable considering the fact that the pitch and the VAE latent feature are obtained from the waveform separately (\textit{tail-to-tail} connection in the graphical model).}. Therefore, $q(z,y|x)$ can be factorized as $q(z|x)q(y|x)$.

Furthermore, as pitch extraction from $x$ is a deterministic operation, we can define $q(y|x)=\delta(y-y_{gt})$ and transform the third term of \eqref{eq:elbo} as follows:
\begin{equation}
    - \log p(y_{gt}|c) + const. ,
\end{equation}
where $y_{gt}$ represents the observed ground truth pitch value. We can optimize this part by minimizing the L2 norm between predicted and ground truth by assuming a Gaussian distribution for $p(y|c)$ \sedit{with fixed unit variance}.
The three terms in \eqref{eq:elbo} can then be interpreted as wave reconstruction loss of VAE $L_{recon}$, KL divergence loss between prior/posterior distributions $L_{kl}$, and pitch reconstruction loss from text $L_{pitch}$, respectively. Following~\cite{kim2021conditional}, we adopt mel-spectrogram loss for $L_{recon}$.

\subsection{Prior encoder}
As the proposed method not only focuses on reading-style TTS, but also TTS with an emotional dataset with significant diversity in terms of pronunciation, the prior distribution modeled by the prior encoder needs to represent the rich acoustic variation of pronunciation within the same phoneme.
To this end, we adopt the frame prior network proposed in~\cite{zhang2022visinger}. It expands phoneme-level prior distribution to frame-level fine-grained distribution. We confirmed in preliminary experiments that this was effective to stabilize the pronunciation not only for singing voice synthesis but also for multi-speaker emotional TTS. In addition, we introduce a frame pitch predictor from a hidden layer of the frame prior network to predict the frame-level prosodic features, i.e., fundamental frequency ($F_0$) and voicing flag ($v$), which are subsequently used as inputs for the periodicity generator described in \secref{subsec:pgen}. As discussed in \secref{subsec:model_overview}, these features are optimized using the L2 norm, as follows\footnote{\sedit{In our preliminary experiment, we also investigated cross-entropy loss for $v$ instead of L2 loss and found that both criteria performed equally well.}}:
\begin{equation}
    L_{pitch} = \|\log F_0 - \log \hat{F_0}\|_2 + \| v - \hat{v}\|_2.
\end{equation}
The prior distribution is augmented by normalizing flow $f$ to enhance the modeling capability, as in VITS:
\begin{equation}
    p(z|c) = N(f(z); \mu(c), \sigma(c))\left|\det \frac{\partial f(z)}{\partial z}\right|,
\end{equation}
where $\mu(c)$ and $\sigma(c)$ represent trainable mean and variance parameters calculated from text representation, respectively.

\subsection{Decoder with periodicity generator}
\label{subsec:pgen}
\label{subsec:generator}
\begin{figure}[t]
    \centering
    \includegraphics[width=7cm]{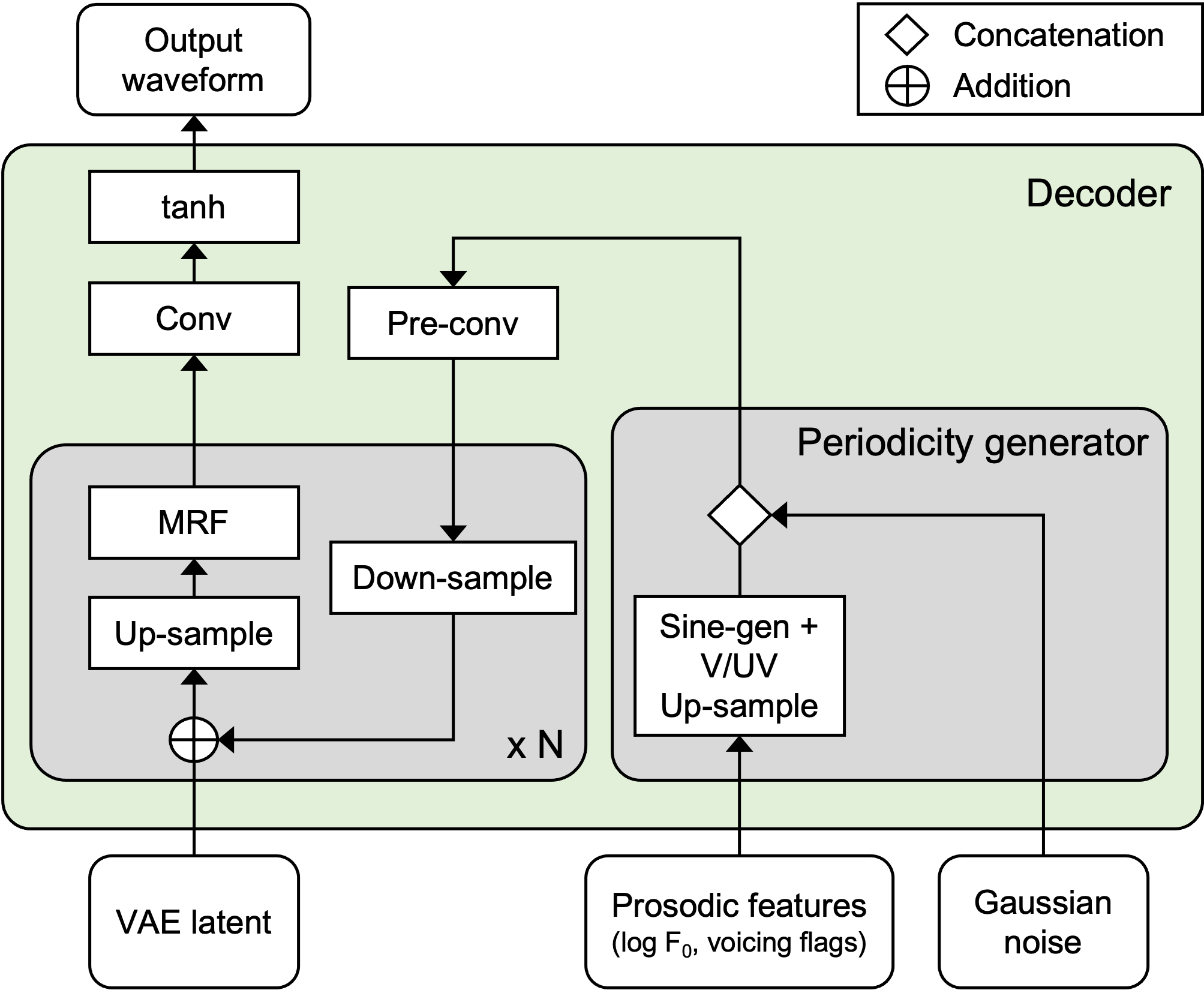}
    \caption{\small Decoder architecture with periodicity generator. MRF denotes multi-receptive field fusion in HiFi-GAN~\cite{kong2020hifi}. Up/down-sampling is performed by transposed and normal 1-D convolution, respectively. }
    \label{fig:generator}
\end{figure}
It has been reported that GAN-based vocoder models typically produce artifacts when reconstructing waveforms from acoustic features due to their inability to estimate pitch and periodicity~\cite{morrison2021chunked}. We found that these artifacts are also observed in end-to-end TTS models, particularly when trained on a dataset with large pitch variance, such as an emotional one.
To address this problem, we use a sine-based source signal to explicitly model the periodic component of speech waveforms, which has proven to be effective in some previous works~\cite{hono2021periodnet,hwang2021high,wang2019neural}.
However, it is not straightforward to incorporate it into the HiFi-GAN-based decoder (i.e., vocoder) architecture in VITS, as a sine-based source signal is supposed to be a sample-level feature, while the input of HiFi-GAN is typically a frame-level acoustic feature\footnote{We investigated other decoder architectures, such as that in~\cite{hwang2021high}, but they did not work well with end-to-end optimization.}.

To overcome this mismatch, we devise a model architecture inspired by a pitch-controllable HiFi-GAN-based vocoder~\cite{matsubara2022phg}.
Figure~\ref{fig:generator} shows the decoder architecture of the proposed model. The key idea is to successively apply down-sampling layers to the sample-level pitch-related input to match the resolution of the up-sampled frame-level feature.
We call the module to generate a sample-level periodic source as periodicity generator.
We input the sinusoidal source together with the voicing flags and Gaussian noise, as this setting performed better in a previous work~\cite{hwang2021high}.
In addition, unlike the previous work in~\cite{matsubara2022phg}, we avoid directly adding the sample-level output of the pre-conv module in \figref{fig:generator} to the up-sampled features, as we found this degrades the sample quality.

\subsection{Training criteria}
\label{ssec:criteria}
In addition to the aforementioned losses, we adopt adversarial loss $L_{adv}$ and feature matching loss $L_{fm}$ to train a generative adversarial network (GAN) for waveforms~\cite{mao2017least,larsen2016autoencoding}. L2 duration loss $L_{dur}$ is also employed to train a supervised duration model.
Note that the details of the GAN-related loss terms are described in~\cite{kim2021conditional}.
The total loss $L_{total}$ for training is then summarized as follows:
\begin{equation}
    L_{total} =  L_{recon} + L_{kl} + L_{pitch} + L_{dur} + L_{adv} + L_{fm}.
\end{equation}

\section{EXPERIMENTS}
\label{sec:experiments}
\subsection{Experimental setup}
\label{ssec:exp_setup}
\subsubsection{Database and feature extraction settings}
For the experiments, we used phonetically-balanced emotional speech corpora recorded by five male and ten female Japanese professional speakers.
We sampled speech signals at 24~kHz with 16~bit quantization.
All the 15 speakers' data contained three speaking styles: \textit{neutral}, \textit{happy}, and \textit{sad}. The number of utterances for each style was 4000, 1000, and 1000, for each speaker, respectively. For every speaker--style pair, 50 utterances were kept for each validation and test, respectively. The rest were used for training, which amounted to 84.08 hours.
We extracted a $513$-dimensional linear-spectrogram for end-to-end models and a $80$-dimensional log-mel spectrogram for cascade models with a $10$~ms frame shift and a 40 ms window length. We also extracted continuous log $F_0$ and voicing flags using the improved time-frequency trajectory excitation vocoder~\cite{yu2010continuous,song2017effective}.
Acoustic features for the cascade models were constructed by concatenating these pitch features to the log-mel spectrogram.
The phoneme durations to train the supervised duration model were manually labeled by professional annotators.
We normalized the acoustic features so that they had a zero mean and unit variance for each dimension using the statistics of the training data.
\subsubsection{Model details}

\begin{table*}[t]
    \vspace{-2mm}
    \caption{
    \fontsize{9.0}{10.6}\selectfont
    Comparison of the systems. E2E: end-to-end model. FPN: frame prior network in VISinger~\cite{zhang2022visinger}. FS2: FastSpeech~2~\cite{ren2020fastspeech}. P-VITS: Period VITS (i.e., our proposed model). V/UV: voicing flags. *: Not the same but a similar architecture.}
    \vspace{1mm}
    \centering
    \scalebox{0.92}{
    \begin{tabular}{llllll} \hline
       Model  & Type  & Duration input & FPN & Pitch input & Periodicity generator input \\\hline
       VITS     & E2E & No & No & None & - \\
       FPN-VITS & E2E & Yes & Yes & None & - \\
       CAT-P-VITS & E2E & Yes & Yes & Frame-level to decoder & - \\
       Sine-P-VITS & E2E & Yes & Yes & Sample-level to decoder & Sine-wave \\
       P-VITS (proposed) & E2E & Yes & Yes & Sample-level to decoder & Sine-wave + V/UV + noise \\
       FS2+P-HiFi-GAN & Cascade & Yes & * & Sample-level to decoder & Sine-wave + V/UV + noise \\ \hline
    \end{tabular}
    }
    \label{tab:systems}
\end{table*}

The training configurations of the proposed method basically followed those of the original VITS~\cite{kim2021conditional}. Regarding the loss criteria in \secref{ssec:criteria}, we set the weights of $L_{recon}$, $L_{kl}$, $L_{pitch}$, $L_{dur}$, $L_{adv}$ and $L_{fm}$ empirically to 45, 1, 1, 1, 1, and 2, respectively.
As our model is a multi-speaker emotional TTS model, we used speaker and emotion embeddings as the global conditions.
In addition, we used accent information as external input to synthesize speech with natural prosody for the Japanese language~\cite{yasuda2019investigation}.

The frame prior network was composed of six 1-D residual convolutional stacks with a kernel size of 17. The frame pitch predictor was composed of a stack of five 1-D convolutions with a kernel size of 5 and a dropout rate of 0.3. Conditional embedding was also added to predict the speaker-dependent pitch contour.
The sine-wave from $F_0$ was generated using an open-source implementation of neural source-filter models\footnote{\url{https://github.com/nii-yamagishilab/project-NN-Pytorch-scripts}}~\cite{wang2019neural_aslp}.
Voicing flags in the periodicity generator were up-sampled from the frame level to the sample level using nearest neighbor up-sampling. We used up-sample rates $[6,5,2,2,2]$ for the decoder, as one frame contains 240 samples.
During training, we used a dynamic batch size with an average of 26 samples to create a minibatch~\cite{hayashi2020espnet}.

Baseline systems are summarized in~\tabref{tab:systems}. We prepared several variants of VITS;
\textbf{VITS} is the original VITS;
\textbf{FPN-VITS} is a model that introduces a frame prior network to VITS; and \textbf{CAT-P-VITS} is similar to \textbf{FPN-VITS}, but additionally concatenated the frame-level $F_0$ and voicing flags to the latent variable $z$. Note that the periodicity generator was not used in this model.
\textbf{Sine-P-VITS} was similarly configured to the proposed method \textbf{P-VITS}, but it omitted voicing flags and Gaussian noise from the input of the periodicity generator.

We also prepared the cascade-type TTS model \textbf{FS2+P-HiFi-GAN}, as the baseline system.
Specifically, we adopted a FastSpeech~2-based acoustic model~\cite{ren2020fastspeech} and a HiFi-GAN-based vocoder~\cite{kong2020hifi}.
The acoustic model used six Transformer layers and six lightweight convolution blocks to compose the encoder and decoder, respectively.
Note that the lightweight convolution blocks perform comparably to or better than the Transformer ones in the TTS context~\cite{elias2021parallel}.
For each block, we set the hidden sizes of the self-attention and convolution layers to 384.
The hidden size of feed-forward layers in Transformer was 768. To bring the condition closer to end-to-end models, pitch and energy predictors in a variance adaptor were omitted.
The vocoder architecture was the same as that of the proposed model.
We trained the acoustic model for 200 K steps using the RAdam optimizer~\cite{liu2019radam} and trained the vocoder for 800 K steps using the AdamW optimizer~\cite{loshchilov2017decoupled}. We further fine-tuned the vocoder by using the output of the acoustic model for 400 K steps to improve the quality.

\subsection{Evaluation}
To evaluate the effectiveness of the proposed method, we conducted a subjective listening test using a 5-point naturalness mean opinion score (MOS). We asked 18 native Japanese raters to make a quality judgment.
We randomly selected 4 speakers for the test and picked up 15 utterances from the test set for each speaker--emotion pair. In total, the number of utterances for each system was 15 utterances * 4 speakers * 3 emotions = 180.
We evaluated the recorded speech and synthetic speech from the six TTS systems shown in \tabref{tab:systems}.

\label{sec:results}
\begin{table}[t]
    \vspace{-3mm}
    \caption{
    \fontsize{9.0}{10.6}\selectfont
    Naturalness MOS test results with 95\% confidence interval.
    \textbf{Bold font} represents the best scores except for the reference.
    }
    \vspace{1mm}
    \centering
    \scalebox{0.92}{
    \begin{tabular}{llll} \hline
       Model  & Neutral & Happy & Sad \\\hline
       Reference & 4.66 ± 0.04 &4.77 ± 0.04 &4.71 ± 0.04 \\
       VITS     & 2.78 ± 0.07 &2.82 ± 0.08 & 3.24 ± 0.08\\
       FPN-VITS & 3.85 ± 0.07 &3.69 ± 0.07 &3.77 ± 0.07\\
       CAT-P-VITS & 3.79 ± 0.07& 3.63 ± 0.07 &3.66 ± 0.07\\
       Sine-P-VITS &4.63 ± 0.04 &4.54 ± 0.04 &4.68 ± 0.04\\
       P-VITS (proposed) & \textbf{4.66 ± 0.04} & \textbf{4.62 ± 0.04} & \textbf{4.69 ± 0.04}\\
       FS2+P-HiFi-GAN &4.50 ± 0.05 &4.18 ± 0.06 &4.40 ± 0.05\\\hline
    \end{tabular}}
    \label{tab:mos}
    \vspace{-3mm}
\end{table}

\tabref{tab:mos} shows the results of the MOS evaluation. We can see that the proposed \textbf{P-VITS} performed the best among all the candidate models in terms of naturalness.
Furthermore, our method achieved comparable scores to the reference except for the happy style with no statistically significant difference in student's $t$-test with a 5~\% significance level.
Other notable findings can be summarized as follows. 1) Just concatenating the frame-level pitch feature to the VAE latent feature $z$ was insufficient.
The sample quality improved only when the pitch feature was up-sampled to the sample level by the periodicity generator (\textbf{P-VITS} vs. \textbf{CAT-P-VITS}, \textbf{CAT-P-VITS} vs. \textbf{FPN-VITS}).
2) Although a cascade model with a periodicity generator performed well,
we successfully improved the quality by optimizing it in an end-to-end manner.
The quality gap appeared the most in happy style, where the prosodic variation is the largest (\textbf{P-VITS} vs. \textbf{FS2+P-HiFi-GAN}).
3) The augmentation of prior distribution by frame prior network was also effective for multi-speaker emotional TTS (\textbf{FPN-VITS} vs. \textbf{VITS}).
4) Noise and voicing flag input to the periodicity generator further improved the sample quality (\textbf{P-VITS} vs. \textbf{Sine-P-VITS}).

\begin{figure}[t]
    \begin{minipage}[b]{0.32\linewidth}
        \centering
   	    \includegraphics[width=1.0\linewidth]{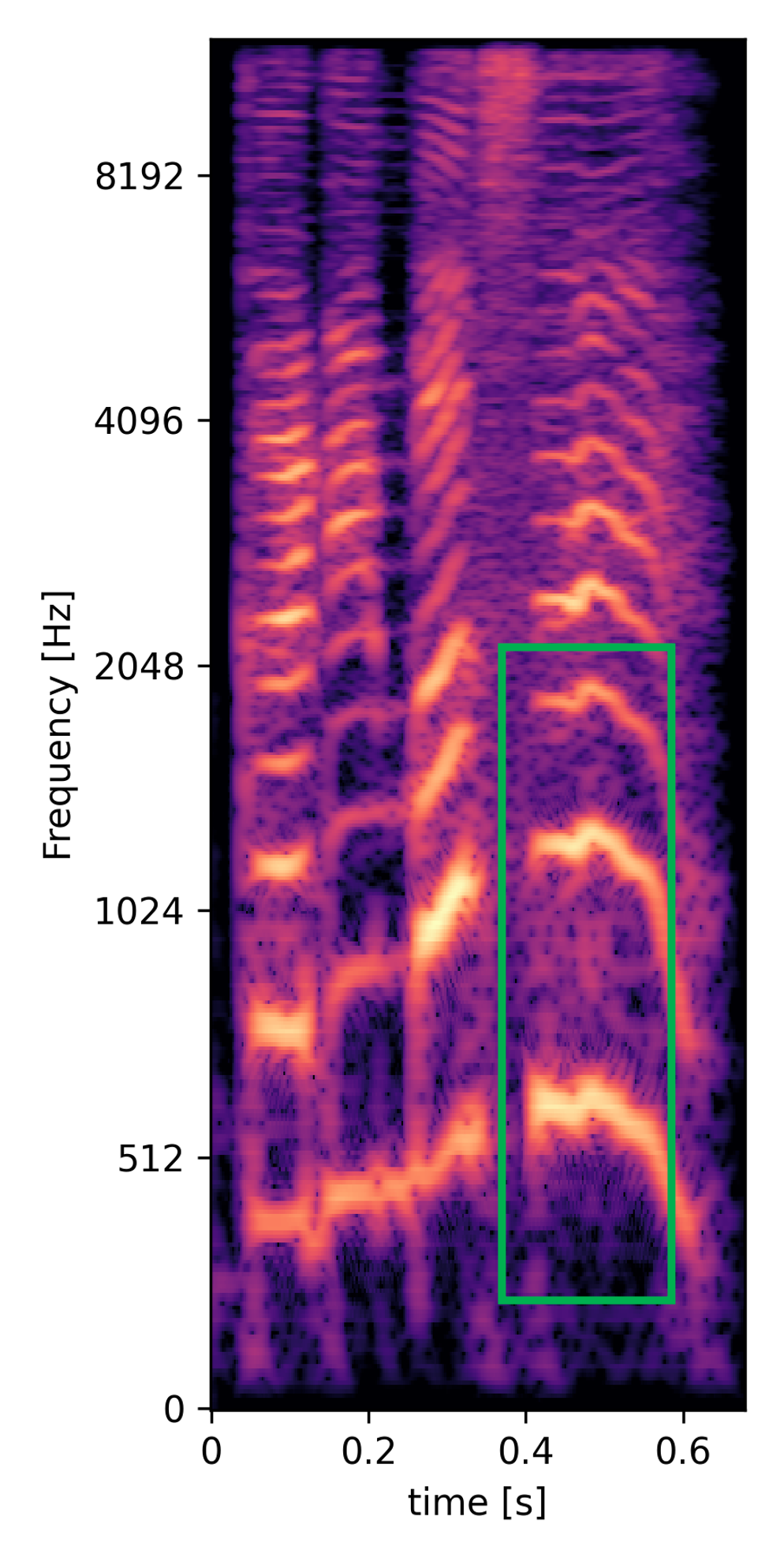}
   		\centerline{\small \textbf{(a) CAT-P-VITS}}  \medskip
        \vspace{-7mm}
    \end{minipage}
   	\begin{minipage}[b]{0.32\linewidth}
        \centering
   	    \includegraphics[width=1.0\linewidth]{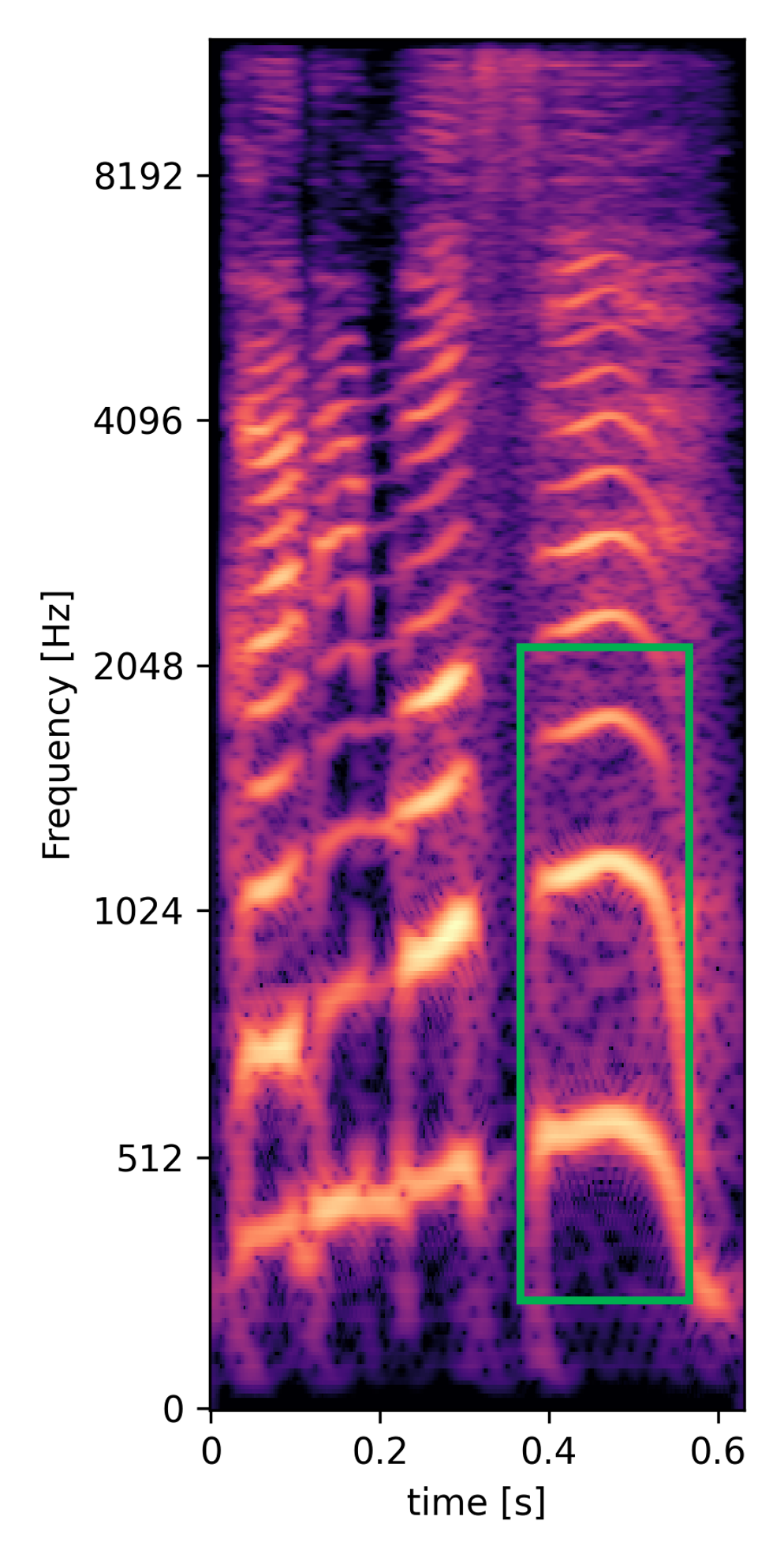}
   		\centerline{\small \textbf{(b) P-VITS}}  \medskip
        \vspace{-7mm}
    \end{minipage}
    \begin{minipage}[b]{0.32\linewidth}
        \centering
   	    \includegraphics[width=1.0\linewidth]{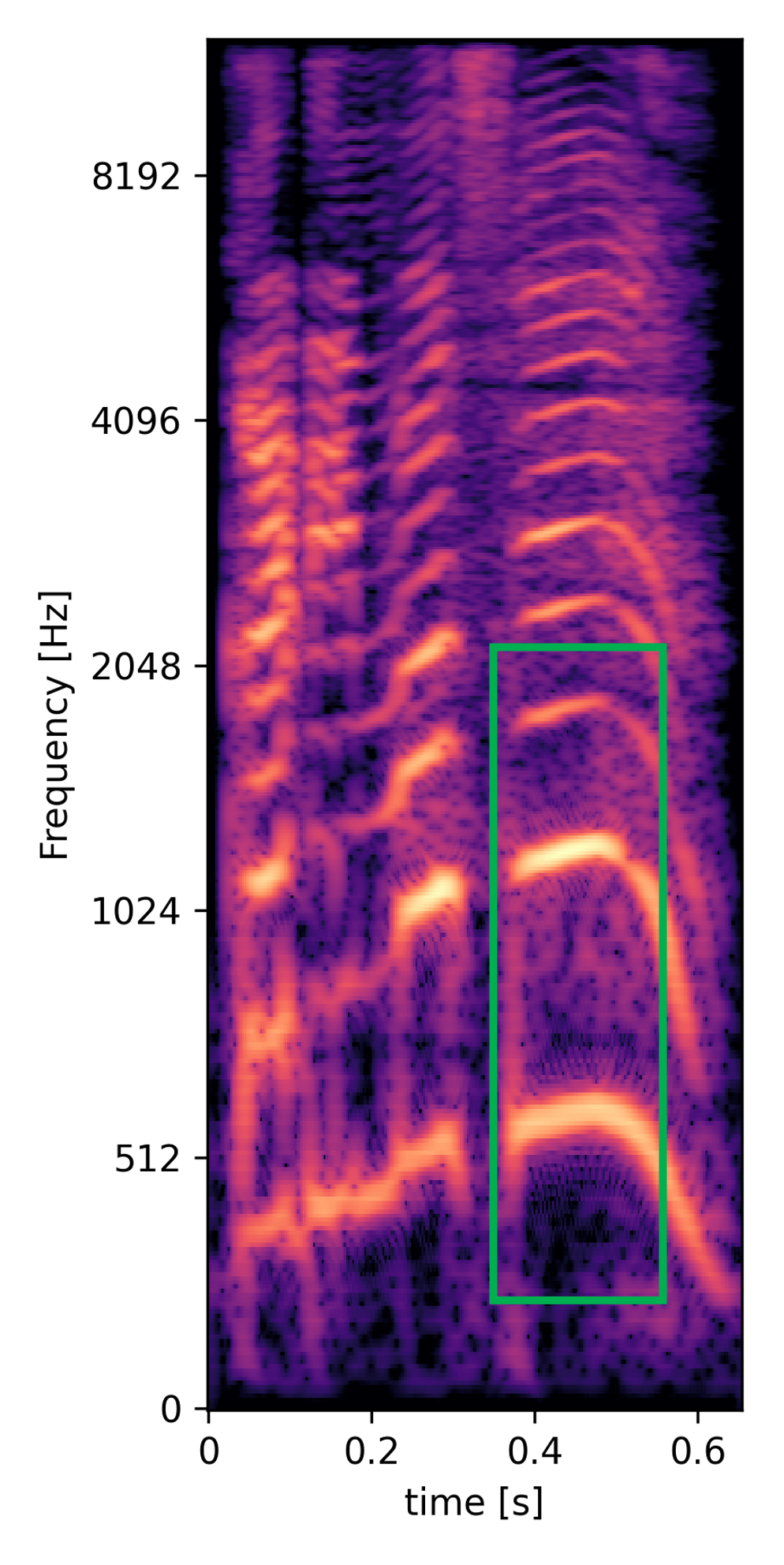}
   		\centerline{\small \textbf{(c) Reference}}  \medskip
        \vspace{-7mm}
   	\end{minipage}
    \caption{
    \fontsize{9.0}{10.6}\selectfont
    Mel-spectrograms of the synthesized and natural speech. The stability of the harmonic structure in the green boxes shows a clear difference between \textbf{(a)} and \textbf{(b)}.}
    \label{fig:spec}
    \vspace{-2mm}
\end{figure}

To confirm the improvement in the pitch contour, we also visualized the mel-spectrogram of the synthesized speech in \figref{fig:spec}. Note that \textbf{CAT-P-VITS} was chosen here to represent the baseline models without periodicity generator.
We can see a swaying pitch contour in \textbf{(a)} at the end of the utterance (green box in the figure), while it is continuous in the proposed method \textbf{(b)}, which is the same tendency to that in the reference \textbf{(c)}.
We encourage readers to listen to the samples provided on our demo page\footnote{\url{https://yshira116.github.io/period_vits_demo/}}.
\vspace{-2mm}
\section{CONCLUSIONS}
\label{sec:con}
We proposed Period VITS, a fully end-to-end TTS system with explicit pitch modeling capable of high-quality speech synthesis even when the dataset contained significant prosodic diversity.
Specifically, the proposed method introduced a periodicity generator that generated sample-level pitch representation from extracted frame-level pitch. The representation was successively down-sampled and added to the HiFi-GAN-based \sedit{decoder} part to stabilize the pitch of the target waveform. The proposed model was optimized in the end-to-end schema from a variational inference point of view. The experimental results showed that our proposed method outperformed the baseline methods in terms of emotional TTS.
Future work should include implicit pitch modeling to avoid pitch extraction errors and to improve the sample quality for highly expressive styles such as the happy style. 
\vspace{-2mm}
\section{Acknowledgments}
This work was supported by Clova Voice, NAVER Corp., Seongnam, Korea.

\vfill\pagebreak

\bibliographystyle{myIEEEbib}
{
\fontsize{9.0}{10.6}\selectfont
\bibliography{strings,refs}
}

\end{document}